
\magnification=\magstep1
\def\ltorder{\mathrel{\raise.3ex\hbox{$<$}\mkern-14mu
             \lower0.6ex\hbox{$\sim$}}}

{}
\vskip 2in
\centerline{\bf Competition Between Pressure and Gravity Confinement in
Ly$\alpha$ Forest Observations}
\bigskip
\bigskip

\centerline{{\bf Jane C. Charlton}\footnote{$^1$}{Astronomy and Astrophysics
Department, Davey Laboratory, Pennsylvania State University, University Park,
PA 16802}$^,$\footnote{$^2$}{Center for Gravitational Physics and Geometry,
Pennsylvania State University}{\bf , Edwin E. Salpeter}\footnote{$^3$}{Space
Sciences Building, Cornell Univ., Ithaca, NY 14853} {\bf , and Suzanne M.
Linder{$^1$}}}

\def\msun{\hbox{$\hbox{M}_{\odot}$}}

\vskip 1in
\leftline{\bf Abstract}
\bigskip

A break in the distribution function of Ly$\alpha$ clouds (at
a typical redshift of $2.5$) has been reported by Petitjean
et al. (1993).  This feature is what would be expected from
a transition between pressure confinement and gravity confinement
(as predicted in Charlton, Salpeter, and Hogan (1993)).  The
column density at which the feature occurs has been used to
determine the external confining pressure, $\sim 10 {\rm cm}^{-3}
{\rm K}$, which could be due to a hot, intergalactic medium.
For models that provide a good fit to the data, the contribution
of the gas in clouds to $\Omega$ is small.  The specific shape
of the distribution function at the transition (predicted by
models to have a non-monotonic slope) can serve as a diagnostic
of the distribution of dark matter around Ly$\alpha$ forest
clouds, and the present data already eliminate certain models.

\bigskip
\medskip
\leftline{Subject headings: quasars: absorption lines - intergalactic medium -
dark matter}
\bigskip
\leftline{Submitted to the {\sl Astrophysical Journal Letters}}

\vfill\eject

\leftline{\bf 1. Introduction}
\medskip

Clouds containing neutral hydrogen (with column densities ranging
from $10^{13} {\rm cm}^{-2}$ up to $10^{22} {\rm cm}^{-2}$) are
observed through Ly$\alpha$ absorption in quasar spectra.
In this Letter we restrict ourselves to the
epoch with the most reliable data, redshift $z \sim 2$ to $3$.
The distribution function $f(N_{Hobs})$ for observed HI
column density of individual Ly$\alpha$ absorption lines is
an important clue to the nature of the absorbing clouds, and
the accuracy of determinations of $f$ is improving steadily.
With little loss of generality, an individual ``Ly$\alpha$
forest cloud'' with some $N_{Hobs}$ can be modeled as a
portion of a slab in hydrostatic equilibrium with
perpendicular HI column density $N_H = N_{Hobs} \cos \theta$
(where $\theta$ is the inclination angle) and an (unknown)
combined HI and HII column density $N_{tot}$ (Charlton, Salpeter,
\& Hogan 1993 (CSH)).  This modeling, in terms of a few
adjustable parameters, is similar whether there are many
different types of individual small clouds (minihalos), each with
a different characteristic value of $N_H$, or a single
type of very extended disk structure (large proto-galaxies)
with $N_H$ decreasing enormously from the center outwards.

Any realistic modeling of ``forest clouds'' must include
the effects of both external pressure and self-gravity.
CSH predicted a break in the distribution function, $f(N_{Hobs})$,
where these two effects are equal.  There must be a transition
between pressure and gravity domination for some value
$N_{H1obs}$ of column density.  The most recent observational
data (see Petitjean et al. 1993 (PWRCL)) shows a break of the
type predicted, with $N_{H1obs} \sim 10^{16} {\rm cm}^{-2}$.
The transition value predicted by the CSH models depends on
the value of the external pressure, $P_{ext}$, and weakly
on the ratio of dark matter to gas density at the transition;
thus we shall be able to estimate $P_{ext}$.

In this Letter we constrain model parameters by fitting
to the neutral column density distribution over the forest
regime.  Since the observed distribution function $f(N_{Hobs})$ is of
piece-wise power law form, it is a reasonable assumption
that the basic distribution function $g$ of the total
(perpendicular) column density $N_{tot}$ is a continuous
power law.  The observed $f(N_{Hobs})$ will be fit by
adjusting three parameters: 1) the value of $N_{H1}$ for
the transition between pressure and gravity dominated regimes (set
by the external pressure $P_{ext}$ and a dark matter
parameter $\eta$); 2) the slope of an assumed power
law dependence of $\eta$ on $N_{tot}$ (The pressure
induced by the self-gravity of the gas alone is proportional
to $N_{tot}^2$, but the gravity due to dark matter present
in the slab increases this pressure by a factor of $\eta$.)
;  3) the slope of
the power law $g(N_{tot})$ distribution.  We shall examine
all of this three-dimensional parameter space, but give the
specific example of extended proto-galactic disks in
Section 3c.

Alternative models for the observed ``break'' have been
given by PWRCL in terms of four types
of clouds and a transition from metal-poor to metal-rich
systems, but these require more parameters than the
CSH model.  We also omit, in this paper, possible
deviations from ionization and thermal equilibrium
(Miralda-Escud\'e and Rees 1993), since we wish to
test the simplest models first.

Section 2 uses the relationship
between neutral and total column density in various
regimes (pressure dominated, gravity dominated, Lyman limit,
and damped) to schematically
illustrate the various features in the
distribution function of $N_H$.  Section 3 describes more
detailed models that include corrections for orientation,
extinction at the upper end of the forest regime, and
consideration of the vertical slab structure.  The best fit
model parameters (and plausible ranges) to the PWRCL
column density distribution are presented.
Also, the number of systems expected in the Lyman limit
regime and the total contribution of forest
systems to the mass density of the Universe are computed.
Section 4 concludes with a summary
of the constraints on the parameters.  A value of $P_{ext}$
is derived, and the implications of constraints on the
dark matter distribution are discussed.

\bigskip
\leftline {\bf 2. Overview of the Different Regimes}
\medskip

The distribution $g(N_{tot})$ is assumed to have a power
law form.  Written in logarithmic notation,
$g'(N_{tot}) = N_{tot} g(N_{tot})$, and we define the parameter
$\epsilon'$ by
$$
g'(N_{tot}) d\ln N_{tot} = C N_{tot}^{-\epsilon'} d\ln N_{tot} . \eqno(1)
$$
The distribution function $f'(N_H) = N_H f(N_H)$ for
neutral hydrogen, in logarithmic notation, can be derived
if $N_H$ is given as a function of $N_{tot}$.  Since
$f'(N_H) d\ln N_H = g'(N_{tot}) d\ln N_{tot}$ we have
$$
f'(N_H) = g'(N_{tot})/k(N_{tot}) ,
$$
$$
k(N_{tot}) = d\ln N_H/d\ln N_{tot} . \eqno(2)
$$

To derive $N_H$ as a function of $N_{tot}$
in the forest regime,
we model Ly$\alpha$ clouds as slabs in hydrostatic equilibrium
and ionization/recombination balance.
(For simplicity we neglect the presence of He.)  The basic
differences between behavior in the pressure and gravity
dominated regimes can be illustrated by balancing the forces
in a half slab approximation:
$$
{\pi \over 2} G {m_H}^2 \eta N_{tot}^2 + P_{ext} = 2 n_{tot} kT \eqno(3)
$$
where $n_{tot}$ represents the total number density of H (at
a ``typical'' vertical distance from slab center).
On the LHS, the first term is the force per unit
area due to gravity, and $P_{ext}$ is due to the inward,
external force.  The outward force (RHS) is due to gas pressure
within the slab, with the factor of two arising because slabs
are highly ionized so that free electrons also contribute
to this pressure.  The parameter $\eta$ is the enhancement
factor for gravity over the gas column due to dark matter,
taken to depend on $N_{tot}$ as
$$
\eta(N_{tot}) = \eta_1 (N_{tot}/N_{t1})^{-\delta} \eqno(4)
$$
where $\eta_1$ and $N_{t1}$ are values at the transition
between the pressure and gravity dominated regimes (where
the two terms on the LHS of equation (3) are equal).
For the case with no dark matter we have $\eta_1 = 1$ and
$\delta = 0$.  Using
ionization/recombination balance for ionization rate
$\zeta$ (obtained from the proximity effect (Bajtlik, Duncan,
\& Ostriker 1988), a temperature of 20000K, and equating
the gravity and pressure
terms in equation (3) gives $P_{ext}$
in terms of the observable $N_{H1}(\sim N_{H1obs}/1.8)$ as
$$
P_{ext}/k = 14.8 \eta_1^{1/3} \left ( {{\zeta}\over{2.7 \times 10^{-12}
{\rm s}^{-1}}} \times {{N_{H1obs}} \over
{10^{16} {\rm cm}^{-2}}} \right )^{2/3} {\rm cm}^{-3} {\rm K}. \eqno(5)
$$
The relationship between $N_H$ and $N_{tot}$ in
terms of their values at the transition, $N_{H1}$ and $N_{t1}$,
is
$$
\left ( {{N_H} \over {N_{H1}}} \right )_{hs} = 0.5
\left [ \left ( N_{tot} / N_{t1} \right )^{3-\delta}
+ \left ( N_{tot} / N_{t1} \right ) \right ]
\eqno(6)
$$
where the subscript ``hs'' denotes the ``half-slab'' approximation
used in this section.

Of necessity there are four different regimes for the
relationship between $N_H$ and $N_{tot}$ in the range
from the smallest to the largest values of column density,
shown schematically in Figure 1a:
As mentioned, there must be
some value $N_{t1}$ of $N_{tot}$ (with a corresponding
value $N_{H1}$ of $N_H$) where the external and self-gravity
pressures are equal.  Far to the left of this
transition (region (a) in Figure 1) the pressure is
constant ($=P_{ext}$), $N_H$ is proportional to
$N_{tot}$ (eq. (6)) and $k$ (eq. (2))
is approximated by the constant $k_a = 1$.  Far
to the right of the transition (but still with
$N_H << 10^{17.2}$)
the ratio $N_H/N_{tot}$ increases with increasing
$N_{tot}$ as $N_{tot}^{2-\delta}$, so that in this
gravity dominated regime (b) the derivative $k$ is
approximated by the constant $k_b = 3 - \delta$.
The next transition point (2) occurs when $N_H$ is
near the value $N_{H2} = 10^{17.2} {\rm cm}^{-2}$
where the optical depth for a (normally incident)
photon just above the Lyman limit is $1$.  Just
to the left of this transition, i.e. in the
upper range of region (b), $N_H/N_{tot}$ increases
slightly more rapidly than $N_{tot}^{2-\delta}$
because photon extinction decreases the ionizing
flux and hence increases the neutral fraction.
To the right of this transition, over a fairly
narrow range of $N_{tot}$, from $\sim N_{t2}$
to $N_{t3} \sim 2N_{t2}$, lies a region (c) which
displays the ``Stromgren effect'', first
suggested by Sunyaev (1969) and Felten and Bergeron (1969)
and discussed in detail by Maloney (1993) and
Corbelli and Salpeter (1993).  The behavior of
$N_H$ in this region is complicated, but the
essential point is that $N_H$ increases by a
very large factor, from $N_{H2}$ to $N_{H3} \sim
N_{t2} >> N_{H2}$.  Finally, in the fourth region
(d) where $N_{tot} >> N_{t3}$, the gas is almost
entirely neutral, $N_H \approx N_{tot}$ and
$k_d \approx 1$.

If equation (1) held for all values of $N_{tot}$, the
distribution function $f'(N_H)$, given by equation (2),
would also display the four regions shown in Figure 1b.
If we define $\beta ' = \beta - 1$
(where $f(N_H) \propto N_H^{-\beta}$) by
$$
f'(N_H) \propto N_H^{-\beta'}, \eqno(8)
$$
then well
into regions (a) and (d), $N_H \propto N_{tot}$ and
$k_a = k_d = 1$ so that $\beta_a' = \beta_d' = \epsilon'$.
Similarly, well into region (b), equation (8) holds with
$\beta'_b = \epsilon'/k_b$ and $k_b = 3-\delta$.
However, just before the transition from a to b the
function $k(N_{tot})$ increases before settling into
the constant $k_b$, so that equation (2) predicts a
short small dip, i.e. the slope $d\ln f'(N_H)/d\ln N_H$
does not decrease monotonically (see Figure 2a).  The extinction
correction (see section 3) near $N_{H2}$ has a similar
effect of increasing $k$ and there is again an increase
in slope.  In the Stromgren effect region (c) the
extinction increases rapidly and $N_H/N_{tot}$
increases rapidly at first and then approaches unity,
so that $k(N_H)$ increases at first, reaches a maximum
and decreases towards unity.  The actual behavior of
$f'(N_H)$ in region (c), indicated schematically by the
dashed curve in Figure 1b, depends on the shape of the photon
spectrum and on just how large the ratio
$N_{H3}/N_{H2} \sim N_{t2}/N_{H2} >> 1$ is
(see Corbelli and Salpeter 1993), but $f'(N_H)$ is
not much smaller at the high $N_H$ end of region (c) than
at the beginning.  We shall see that $\epsilon' < 1$
for any acceptable model, so that the total gas mass
would diverge on the upper end of $N_{tot}$ if
equation (1) held all the way up.  There must thus be
another transition point (4) beyond which $g'(N_{tot})$
changes and eventually decreases rapidly in region (e).

The extent of the five regions (a) to (e) are not known
a priori, nor are the transition values except for
$N_{H2} \sim 10^{17.2} {\rm cm}^{-2}$ (defined by
approximate unit optical depth at the Lyman edge).
On the observational side there are three known
transition points: $N_H \sim 10^{13} {\rm cm}^{-2}$
for the threshold of Ly$\alpha$ absorption
observations, $N_H \sim N_{H2} \sim 10^{17.2} {\rm cm}^{-2}$
for the onset of the Lyman limit regime, and
$N_H \sim 10^{20} {\rm cm}^{-2}$ for the
onset of the damped wing regime.  Between the last two
points, numbers of absorption lines can be measured,
but not the actual values of $N_H$ and we shall
refer to this range from $N_{Hobs} = 10^{17.68} {\rm cm}^{-2}$
to $N_{Hobs} = 10^{20.5} {\rm cm}^{-2}$ as ``the long bin''
of PWRCL.  In principle, it is possible that
$N_{H1} < 10^{13} {\rm cm}^{-2}$
so that all of the observable Ly$\alpha$ forest regime
would lie in region (b), terminating at $N_{H2} \sim 10^{17.2}
{\rm cm}^{-2}$.  As Figure 1 shows, the predicted slope
in this whole region (from $10^{13} {\rm cm}^{-2}$ to
$10^{17.2} {\rm cm}^{-2}$) would have to be mostly
constant with an increase at the end.  The observational
data of PWRCL suggests a decrease
in slope, at best marginally
compatible with a constant slope and incompatible
with an increase in slope.  We therefore accept
their observational ``break'' near $N_{Hobs} \sim 10^{16}
{\rm cm}^{-2}$ as representing the theoretically
required change in slope at $N_{H1}$ predicted
by CSH.

\bigskip
\leftline{\bf 3. Fitting More Accurate Models}
\medskip

We wish to fit all the observational data in the
``Ly$\alpha$ forest regime'' with
$\log N_{Hobs}$ up to 17.68, using eq. (2) but
adopting several model refinements.

\medskip
\leftline{\bf a) Model Refinements}
\medskip

({\it i}) Instead of the
``half slab approximation'' of section 2, we solve
for the vertical structure of a slab, i.e. the
density $\rho$ as a function of height $z$ (from slab
center).
We assume throughout that dark matter gravity dominates over
that due to the gas, and that the dark matter
density is approximately constant over the height
of the gas slab, so that the gravitational acceleration
$g(z)$ is proportional to $z$.  With this assumption
(plus assuming constant temperature) the shape of
$\rho(z)$ can be calculated in terms of the parameter
$N_{tot}/N_{t1}$ alone (see CSH, Section 5.1).

({\it ii}) In the approximate treatment so far, we have
considered the ionization rate per H atom, $\zeta$, a
constant without allowance for absorption.  By definition,
observers chose for the upper bound to the ``forest regime''
(and lower limit to the ``Lyman limit'' regime, section 3d)
a neutral column density $N_{Hobs}$ where the optical
depth is moderate, $\tau = 3(13.6{\rm eV}/E)^3$ for
$10^{17.68} {\rm cm}^{-2}$).  In the small perturbation
approximation, the percentage increase in $N_H/N_{tot}$
equals the decrease in $\zeta$ due to absorption, which depends only
on $N_H = N_{Hobs} \cos \theta$ and the spectral shape
(we use the usual quasar spectrum).  We have calculated
this small correction ($\ltorder 40$\%) for $N_H$ up to
$10^{17.41}$.

({\it iii}) In reality, different slabs can have any
inclination angle $\theta$.  The distribution of observed
values of $f(N_{Hobs})$ can be derived from the distribution
of perpendicular values (Milgrom 1988).
Note that the orientation effect does not alter the power
law behavior of $f(N_{Hobs})$ far from
$N_{H1}$ (on either side), but is responsible
for a gradual transition instead of a sharp break.

\vfill\eject

\leftline{\bf b) Fits to $N_H$ Distribution}
\medskip

For any choice of the model parameters $N_{H1}$, $\delta$
(eq. (4)), and $\epsilon'$ (eq. (1)), the distribution
$f'(N_H)$ can be computed and normalized to provide the
smallest value of $\chi^2$ when compared to the
PWRCL data.
For various values of the dark matter parameter,
$\delta$, the best fit values of $\epsilon'$ and
$N_{H1}$ are given in Table 1.  The smallest $\chi^2$
value is obtained for $\delta \sim -1.5$, but the
fit is only slightly worse for any $\delta \ltorder 0.65$.
The slope of $f'(N_H)$
as a function of $N_H$ is illustrated in Figure 2a for
various $\delta$, and the $f'(N_H)$ distributions are
plotted along with the data
in Figure 2b.  Using
$g'(N_{tot})$, which we have explicitly for the models in Table 1,
the contribution to the cosmological
density parameter is found to be $\Omega_{for} \sim 10^{-2} \eta_1^{-1/3}$
or slightly less for each of the four cases.

\medskip
\leftline{\bf c) Extended Proto-Galaxy Disks}
\medskip

Hoffman et al. (1993) have proposed that the outer
regions of proto-galaxy disks are responsible for the
Lyman forest lines and that the total column density
$N_{tot}(R)$ has a power law form.  The exponent is
related to $\epsilon '$ in eq. (1) by
$$
N_{tot}(R) \sim R^{-p}, ~~~~~~~~p = 2/\epsilon' . \eqno(8)
$$
If we assume a spherical dark matter halo which gives
a flat rotation curve, we can derive a unique relation
between the parameters $\delta$ and $\epsilon '$:  The
gravitational force per unit area in the GDR (the
first term in eq. (3)) is proportional to
$N_{tot}W/R^2$ where $W=N_{tot}/n_{tot}$ is the
slab width.  Using equation (8) we find
$$
\delta = {1 - 0.5 \epsilon'} , \eqno(9)
$$
Since $\epsilon '$ in Table 1 is close to $0.71$
(p=2.8) this gives $\delta = 0.65$.  If
the rotation velocity declines slowly at large
distances $\delta$ would be smaller.

For assumed values for the constant rotation velocity
$V_c$ and the radius $R_{13}$ where $N_H$ has dropped
to $10^{13} {\rm cm}^{-2}$, one can calculate the
transition dark matter parameter $\eta_1$.  We find
$\eta_1 = 4.7~(V_c/70{\rm km/s})~(R_{13}/250{\rm kpc})$
in terms of values for $V_c$ and $R_{13}$ suggested
by the ``Vanishing Cheshire Cat model'' (Salpeter 1993;
Hoffman et al. 1993).   For these values
$P_{ext}/k = 7 {\rm cm}^{-3}{\rm K}$, and the
Ly$\alpha$ forest
systems come from radii between 24kpc and $R_{13}$, with a
total gas mass of $8 \times 10^{9} \msun$ per
proto-galaxy.  The dark matter
mass (out to $R_{13}$) is $3 \times 10^{11} \msun$ per
proto-galaxy, and the number density of these galaxies
is about 4.2Mpc$^{-3}$ at $z=2.5$ (the spheres would
touch at $z \sim 3.3$).  The contribution to $\Omega$
from the dark matter spheres is $\sim 0.2$.

\medskip
\leftline{\bf d) Model Predictions for Lyman Limit Systems}
\medskip

In section 3b we obtained acceptable model fits to the
observed distribution function $f'(N_{Hobs})$ up to
$\log N_{Hobs} = 17.68$.
In the ``Lyman limit regime''
we presently only have an observational
value $I_{lb} = 0.42$ for the integral of $f$ over a
``long bin'' stretching from
$\log N_{Hobs} = 17.68$ to $\log N_{Hobs} = 20.5$
(PWRCL).  Using equations (1) and (2)
we calculate the integral of $g'$ over the corresponding interval
of $N_{tot}$ and equate it to the observed $I_{lb}$.
For each model in Table 1 we give the dimensionless
ratio $F_{lb} = I_{lb} \epsilon'/k(N_{t2}) f'(N_{H2})$.
If the power law in equation (1) holds for $N_{tot}$ up to
$N_{t3}$, then $F_{lb}$ is given by
$$
F_{lb} = 1 - \left ( N_{t3} / N_{t2} \right )^{-\epsilon'} . \eqno(10)
$$
The inequality $F_{lb} < 1$ (the requirement that the
model predicts the observed number of Lyman limit systems
for a finite value of $N_{t3}$) is violated only for the
model with $\delta = 1.5$

\bigskip
\leftline{\bf IV. Discussion}
\medskip

This letter has focused on fitting equilibrium slab models
of Ly$\alpha$ forest clouds to the neutral column density
distribution of Petitjean et al. (1993) for clouds at
$z \sim 2.5$.  We find that the best fit is provided
by models that have a transition between pressure and
gravity confinement at neutral column densities of
$N_{H1} \sim 10^{15} {\rm cm}^{-2}$.  The best fit
for this $N_{H1}$ yields a $\chi^2$ of 1.78 for
$\epsilon' = .70$ and $\delta = -1.5$.  However, fits
are nearly as good for a larger range of
parameters.  Roughly, we have $14.8 < \log N_{H1} < 15.5$,
$.6 < \epsilon' < .8$, and $\delta < 1$.
In comparison, the best fit model in which the clouds are
all gravity confined ($N_{H1} < 10^{13} {\rm cm}^{-2}$)
has $\epsilon' = 1.3$, $\delta = .7$ and $\chi^2 = 3.5$.
Thus we maintain that the data requires a transition
from pressure to gravity confinement.

Since this transition is found to be at
$N_{H1obs} \sim 2 \times 10^{15} {\rm cm}^{-2}$, equation (6)
yields $P_{ext}/k = 5.1 \eta_1^{1/3} {\rm cm}^{-3} {\rm K}$.
This pressure could be due to
a diffuse, universal, hot intergalactic medium.
The measured value is smaller than limits placed by
the COBE y parameter (distortion of the microwave
background expected from Compton scattering by the IGM
electrons) (Mather, et al. 1994).  CSH favored a larger
value of $P_{ext}$ based on extrapolating a single power
law $g'(N_{tot})$ to the damped wing systems, but
region (e) in Figure 1 need not satisfy this power
law.

Models with $\delta > 1$ give
a bad fit to the Ly$\alpha$ forest systems and
underpredict the number of Lyman limit systems ($F_{lb} > 1$).
In any case, the simplest models for the dark matter distribution
suggest $0 < \delta < 1$ and the proto-galaxy
disk model (with constant rotation velocity and
a spherical halo) gives $\delta \sim 0.6$.
However, a disk galaxy with a declining
rotation velocity and/or a compressed spheroid
for the inner dark matter halo would have an
appreciably smaller value for $\delta$.

Although we have considered only simple models
in this Letter, we hope that two features of
our results are fairly universal: 1) The
external pressure $P_{ext}$ is likely to be
large enough to lead to a pressure/gravity
transition at $\log N_{H1} \sim 15$ and the
neutral fraction of the hydrogen gas is
likely to be large enough so that the gas
in Ly$\alpha$ forest systems contributes
rather little to the cosmological mass
density.  2) Because of the form of equation (2),
the slope of the distribution function
$f'(N_H)$ should be a non-monotonic function
(see Figure 2a).  The present observational
data in Figure 2b already seem to corroborate this
feature.

We acknowledge support from NASA grant
NAGW-3571 at Penn State and NSF grant AST 91-19475
at Cornell, the careful reading of the referee, Phil
Maloney, and the Aspen Center for Physics.

\vfill\eject
\magnification=\magstep1
\leftline{\bf Table 1: Summary of Model Results}
\bigskip

\halign{\hfil#\hfil&\quad\hfil#\hfil&\quad\hfil#\hfil&\quad\hfil#\hfil
&\quad\hfil#\hfil\cr
$\delta$&$\epsilon'$&$\log N_{H1}$&$\chi^2$&$F_{lb}$\cr
-1.5&$.70^{+.11}_{-.10}$&$15.0^{+.4}_{-.2}$&1.78&.21\cr
0&$.71^{+.09}_{-.08}$&$15.1^{+.4}_{-.2}$&1.95&.37\cr
.65&.71&$15.2^{+.3}_{-.2}$&2.18&.57\cr
1.5&$.69^{+.07}_{-.07}$&$15.4^{+.9}_{-.8}$&3.10&1.24\cr
}

\vfill\eject
\def\bib{\smallskip\noindent\hang}
\hangindent=0.3in
\centerline{\bf References}
\bigskip

\bib
Bajtlik, S., Duncan, R. C., \& Ostriker, J. P. 1988, ApJ, 327, 570
\bib
Charlton, J. C., Salpeter, E. E., \& Hogan, C. J. 1993, ApJ, 402, 493 (CSH)
\bib
Corbelli, E., \& Salpeter, E. E. 1993, ApJ, 419, 104
\bib
Felten, J. E., \& Bergeron, J. 1969, Astrophysics Letters, 4, 155
\bib
Hoffman, G. L., Lu, N. Y., Salpeter, E. E., Farhat, B., Lamphier, C., \& Roos,
T. 1993, AJ, 106, 39
\bib
Maloney, P. 1993, ApJ, 414, 41
\bib
Mather, J. C., et al. 1994, ApJ, 420, 439
\bib
Milgrom, M. 1988, A\&A. 202, L9
\bib
Miralda-Escud\'e, J., \& Rees, M. J. 1993, MNRAS, 260, 617
\bib
Petitjean, P., Webb, J. K., Rauch, M., Carswell, R. F., \& Lanzetta, K. 1993,
MNRAS, 262, 499 (PWRCL)
\bib
Salpeter, E. E. 1993, AJ, 106, 1265
\bib
Sunyaev, R. A. 1969, SvA, 13, 5

\vfill\eject
\centerline{\bf Figure Captions}
\bigskip
Figure 1: A schematic illustration of the regimes of behavior
expected for Ly$\alpha$ clouds.  The relationship between the
total column density $N_{tot}$ and the neutral column density
$N_H$ is given in the top half of the figure, and the
lower portion sketches the resulting distribution function
(the total neutral column density contained in systems
with a given $N_H$).  The five regimes are: a) pressure
dominated, b) gravity dominated, c) ``Stromgren sphere''
transition between optically thin and optically thick
clouds, d) optically thick regime in which clouds are
almost neutral, and e) high column density cut-off.
\medskip
Figure 2: The distribution function for the best fit models.
The solid line represents
the extended proto-galaxy disk model.
a) The slope of the distribution function illustrates
a feature in $f'(N_H)$ that is most pronounced for
large negative values of $\delta$.  b) The distribution
function is shown along with
the data ($1\sigma$ error bars) from PWRCL
for the overall best fit model with
$\delta = -1.5$, and for the proto-galaxy disk model.
\bye